\documentclass[
    aps,
    prd, 
    superscriptaddress,
    showpacs,
    nofootinbib,
    amsmath,
    amssymb,
    longbibliography 
]{revtex4-2}

\usepackage{graphicx}
\usepackage{dcolumn}
\usepackage{bm}
\usepackage{hyperref}
\usepackage{color}
\usepackage{orcidlink}
\usepackage{filecontents}
\usepackage{float}
\usepackage{booktabs}
\usepackage{mathtools}
\usepackage{tikz}
\usetikzlibrary{arrows.meta, positioning, shapes.geometric, calc, fit, backgrounds}

\hypersetup{
    colorlinks=true,
    linkcolor=blue,
    citecolor=blue,
    urlcolor=blue
}

\begin{document}

\title{Extraction of the color dipole amplitude with physics-informed neural networks}
\author{Wei Kou\orcidlink{0000-0002-4152-2150}}
\email{kouwei@impcas.ac.cn}
\affiliation{Institute of Modern Physics, Chinese Academy of Sciences, Lanzhou 730000, Gansu Province, China}
\affiliation{Southern Center for Nuclear Science Theory (SCNT), Institute of Modern Physics, Chinese Academy of Sciences, Huizhou 516000, Guangdong Province, China}
\affiliation{School of Nuclear Science and Technology, University of Chinese Academy of Sciences, Beijing 100049, China}
\affiliation{State Key Laboratory of Heavy Ion Science and Technology, Institute of Modern Physics, Chinese Academy of Sciences, Lanzhou 730000, Gansu Province, China}

\author{Xurong Chen}
\email{xchen@impcas.ac.cn}
\affiliation{Institute of Modern Physics, Chinese Academy of Sciences, Lanzhou 730000, Gansu Province, China}
\affiliation{Southern Center for Nuclear Science Theory (SCNT), Institute of Modern Physics, Chinese Academy of Sciences, Huizhou 516000, Guangdong Province, China}
\affiliation{School of Nuclear Science and Technology, University of Chinese Academy of Sciences, Beijing 100049, China}
\affiliation{State Key Laboratory of Heavy Ion Science and Technology, Institute of Modern Physics, Chinese Academy of Sciences, Lanzhou 730000, Gansu Province, China}

\begin{abstract}
The process-independence of the color dipole amplitude is a cornerstone of high-energy Quantum Chromodynamics (QCD). However, standard phenomenological approaches typically rely on rigid parametric ansatzes and often require ad-hoc geometric adjustments to reconcile inclusive and diffractive measurements. To resolve this tension, we introduce Physics-Informed Neural Networks (PINNs) employing a ``Teacher--Student'' strategy. The physics-based momentum-space Balitsky-Kovchegov evolution dynamics act as the ``Teacher,'' constraining the solution manifold, while the network ``Student'' is refined against inclusive HERA $F_2$ data. This approach extracts a model-independent dipole amplitude without assuming initial states. Strikingly, we demonstrate that this amplitude---without parameter retuning or geometric rescaling---successfully predicts the absolute normalization and kinematic dependence of exclusive $J/\psi$ photoproduction cross-sections. This parameter-free prediction of the saturation dynamics provides promising evidence for the process-independence of the gluon saturation scale and establishes PINNs as a transformative paradigm for uncovering non-perturbative QCD structures.
\end{abstract}

\maketitle

\section{Introduction}
\label{sec:intro}
Unraveling the multi-dimensional gluonic structure of the proton constitutes a central frontier in high-energy nuclear physics. In the small Bjorken-$x$ regime, the rapid proliferation of gluons creates a dense state of matter governed by non-linear recombination—the Color Glass Condensate (CGC) \cite{Gribov:1983ivg,Mueller:1985wy,McLerran:1993ni,Gelis:2010nm}. This transition to saturation is formally encoded in the Balitsky-Kovchegov (BK) equation \cite{Balitsky:1995ub,Kovchegov:1999ua,Kovchegov:1999yj}, which describes the non-linear rapidity evolution of the universal color-dipole scattering amplitude. While the BK equation provides a rigorous theoretical foundation, extracting the universal dipole amplitude directly from experimental data remains a non-trivial inverse problem, crucial for testing the robustness of the saturation picture.

However, reconciling inclusive (DIS) and exclusive (diffractive) measurements within a unified theoretical framework remains a long-standing challenge. Standard phenomenological approaches \cite{Golec-Biernat:1998zce,Iancu:2003ge,Kowalski:2006hc} typically rely on rigid parametric ansatzes and often require ad-hoc geometric adjustments to accommodate diverse datasets \cite{Kowalski:2003hm,Mantysaari:2016jaz}. In parallel, machine learning (ML) has emerged as a transformative paradigm in high-energy physics \cite{Larkoski:2017jix,Guest:2018yhq,Radovic:2018dip,Albertsson:2018maf,Carleo:2019ptp,Bourilkov:2019yoi,Schwartz:2021ftp,Karagiorgi:2021ngt,Boehnlein:2021eym,Shanahan:2022ifi,Yang:2022yfr,Li:2022ozl,He:2023zin,Zhou:2023pti,Zhou:2023tvv,Pang:2024kid,Ma:2023zfj,Luo:2024iwf,Chen:2024epd,OmanaKuttan:2023bnb,Shi:2022vfr,Shi:2022fei,Shi:2021qri,Mansouri:2024uwc,Chen:2024mmd,Chen:2024ckb,Wang:2023poi,Bento:2025agw,Gao:2025dkn,Baihaqi:2025kjd} (see also the review \cite{Aarts:2025gyp} ). Deep learning techniques have revolutionized the unbiased extraction of Parton Distribution Functions (PDFs) \cite{NNPDF:2017mvq,Carrazza:2019mzf}, accelerated calculations in Lattice QCD \cite{Shanahan:2022ifi,Shi:2022vfr}, and reconstructed geometries in chromofields \cite{Kou:2024hzd}. More recently, Physics-Informed Neural Networks (PINNs) \cite{Raissi:2017zsi,karniadakis2021physics} have introduced a paradigm shift by embedding physical laws directly into the optimization landscape. This capability makes PINNs uniquely consistent for solving ill-posed inverse problems in QCD thermodynamics and spectroscopy \cite{Aarts:2025gyp,Kou:2025qsg}, yet their potential to solve non-linear evolution equations in the saturation regime remains largely unexplored.

In this work, we resolve this tension by introducing a Physics-Informed Neural Network (PINN) framework that incorporates the PDE-constrained momentum-space BK evolution dynamics directly into the learning process. Unlike traditional fitting methods, our approach employs a ``Teacher-Student" strategy (inspired by \cite{liu2025automatic}) where the BK equation constrains the solution manifold while the network refines the amplitude against inclusive HERA $F_2$ data \cite{H1:2009pze,H1:2013ktq}. Crucially, we demonstrate that this amplitude—without any parameter retuning or geometric rescaling—successfully predicts the differential cross-sections of exclusive $J/\psi$ photoproduction \cite{ZEUS:2004yeh,H1:2005dtp}. This prediction, without additional tuning of the dipole amplitude, provides compelling evidence for the process-independence of the gluon saturation scale and establishes PINNs as a powerful paradigm for bridging perturbative evolution and non-perturbative hadronic structure.

\section{Theoretical Formalism}
\label{sec:theory}

\subsection{Momentum-Space BK Equation in Diffusion Approximation}
The evolution of the dipole scattering amplitude $N(k, Y)$ in momentum space is governed by the BK equation. In the diffusion approximation, the non-linear integro-differential equation can be simplified to a partial differential equation (PDE). By expanding the BFKL kernel $\chi(\gamma)$ around the saddle point $\gamma = 1/2$, the equation takes the form~\cite{Munier:2003vc,Wang:2022jwh}:
\begin{equation}
    \frac{\partial N(L, Y)}{\partial Y} = \bar{\alpha}_s \left[ \chi(-\partial_L) N - N^2 \right],
    \label{eq:BK_general}
\end{equation}
where $Y = \ln(1/x)$ is the rapidity, $L = \ln(k^2/\Lambda_{\text{QCD}}^2)$ is the logarithmic transverse momentum, and $\bar{\alpha}_s = \alpha_s N_c / \pi$. In this work, following standard phenomenological practices, we adopt a fixed strong coupling constant yielding $\bar{\alpha}_s = 0.191$, and set the QCD scale to $\Lambda_{\text{QCD}} = 0.2$ GeV. In the diffusion limit, the kernel operator is approximated as:
\begin{equation}
    \chi(-\partial_L) \approx A_0 - A_1 \partial_L + A_2 \partial_L^2.
\end{equation}
Consequently, the PDE constraint used in our PINN ($\mathcal{L}_{\text{PDE}}$) is explicitly written as:
\begin{equation}
    \frac{\partial N}{\partial Y} = \bar{\alpha}_s \left( A_0 N - A_1 \frac{\partial N}{\partial L} + A_2 \frac{\partial^2 N}{\partial L^2} - N^2 \right).
    \label{eq:BK_diff}
\end{equation}
The coefficients are derived from the BFKL characteristic function: $A_0 = \chi(1/2) = 4\ln 2$, $A_1 = 0$ (in the symmetric frame), and $A_2 = \chi''(1/2)/2 = 14\zeta(3)$. In this work, we use these standard values to enforce the physical evolution. 

While the diffusion approximation and fixed coupling elegantly capture the traveling-wave nature of geometric scaling, they truncate higher-order integral dynamics necessary for true precision fits. Thus, the extracted amplitude serves as an effective amplitude whose validity is most robust in the small-$x$ ($x \le 10^{-2}$) and moderate $Q^2$ regime.

\subsection{Inclusive Structure Function $F_2$}
To directly utilize the momentum-space output of our PINN, the inclusive proton structure function $F_2(x, Q^2)$ is computed via the convolution of the momentum-space photon wavefunctions and the dipole amplitude. The explicit expression is given by \cite{deSantanaAmaral:2006fe}
\begin{equation}
   F_2(x, Q^2) = \frac{N_c Q^2 R_p^2}{4\pi^2} \int_0^1 dz \int_0^{\infty} \frac{dk}{k} |\tilde{\Psi}_{\gamma^*}(k, z, Q^2)|^2 N(k, Y),
    \label{eq:f2}
\end{equation}
Here, $\left|\tilde{\Psi}_{\gamma^*}\right|^2$ represents the squared photon wave function summed over active quark flavors $f \in \{u, d, s, c\}$. We assign the effective masses for the light quarks as $m_u = m_d = m_s = 0.14$ GeV, and for the heavy charm quark as $m_c = 1.4$ GeV. In momentum space, incorporating massive quarks ($\bar{Q}_f^2 = z(1-z)Q^2 + m_f^2$), it takes the form \cite{deSantanaAmaral:2006fe,Amaral:2020xqv}:
\begin{align}
    \left|\tilde{\Psi}_{\gamma^*}{\left(k^2,z;Q^2\right)}\right|^2 &= \sum_f \left(\frac{4\bar{Q}_f^2}{k^2+4\bar{Q}_f^2}\right)^2 e_f^2 \Bigg\{ 
    \left[z^2+(1-z)^2\right] \nonumber \\
    &\times \left[\frac{4\left(k^2+\bar{Q}_f^2\right)}{\sqrt{k^2\left(k^2+4\bar{Q}_f^2\right)}}\operatorname{arcsinh}\left(\frac{k}{2\bar{Q}_f}\right) + \frac{k^2-2\bar{Q}_f^2}{2\bar{Q}_f^2}\right] \nonumber \\
    &+ \frac{4Q^2z^2(1-z)^2+m_f^2}{\bar{Q}_f^2} \nonumber \\
    &\times \left[\frac{k^2+\bar{Q}_f^2}{\bar{Q}_f^2} - \frac{4\bar{Q}_f^4+2\bar{Q}_f^2k^2+k^4}{\bar{Q}_f^2\sqrt{k^2\left(k^2+4\bar{Q}_f^2\right)}}\operatorname{arcsinh}\left(\frac{k}{2\bar{Q}_f}\right)\right] 
    \Bigg\}.
\end{align}
This formulation allows for a direct evaluation of inclusive observables from the momentum-space solution $N(k, Y)$ without intermediate coordinate-space transformations during the training phase.

\subsection{Exclusive Vector Meson Production}
To validate the universality of the extracted amplitude, we calculate the exclusive production of $J/\psi$ mesons. Following Ref.~\cite{Wang:2022jwh}, the scattering amplitude for $\gamma^* p \to V p$ is given by \cite{Watt:2007nr}:
\begin{equation}
    \mathcal{A}_{T,L}(x, Q^2, \Delta) = i \int d^2\mathbf{r} \int_0^1 dz (\Psi_V^* \Psi_{\gamma^*})_{T,L} \cdot \sigma_{\text{dip}}(r, x) \cdot e^{-i\mathbf{\Delta} \cdot \mathbf{b}}.
    \label{eq:amppp}
\end{equation}
Assuming a factorized dependence on the impact parameter $b$, the integration over $b$ yields the diffractive slope factor $e^{-B_D |t|/2}$. Here, the diffractive slope parameter is adopted from the energy-dependent empirical parametrization $B_D(W) = B_0 + 4\alpha^\prime \ln(W/W_0)$ extracted from HERA measurements \cite{ZEUS:2004yeh,H1:2005dtp}, which accounts for the transverse spatial profile of the target. It is worth noting that while the factorized impact-parameter profile $N(r, b) = T(b)N(r)$ does not strictly preserve the local unitarity bound at $b=0$ for arbitrarily large dipoles, it remains a safe and widely adopted phenomenological approximation \cite{Kowalski:2003hm}. In our calculations, the massive charm quark ($m_c \approx 1.4$ GeV) of the $J/\psi$ meson effectively suppresses the large-$r$ contributions ($r \gtrsim 2 \text{ GeV}^{-1}$), keeping the amplitude well below the unitarity limit within the physically relevant integration domain. For this exclusive process, we map the PINN's momentum-space output to coordinate space via the Hankel transform:
\begin{equation}
    \sigma_{\text{dip}}(r, x) = 2\pi R_p^2 \int_0^\infty k dk J_0(kr) N(k, Y).
    \label{eq:hankel}
\end{equation}

The differential cross-section is expressed as \cite{Kowalski:2006hc,Kowalski:2003hm}:
\begin{equation}
    \frac{d\sigma}{dt} = \frac{1}{16\pi} (1+\beta^2) R_g^2 \left( |\mathcal{A}_T|^2 + |\mathcal{A}_L|^2 \right),
    \label{eq:dsdt}
\end{equation}
where $\beta$ accounts for the real part of the scattering amplitude and $R_g$ is the skewedness correction factor arising from the difference between diagonal and off-diagonal gluon distributions. Assuming a local power-law dependence of the amplitude $\mathcal{A} \propto (1/x)^\lambda$, these factors are analytic functions of $\lambda$ \cite{Shuvaev:1999ce}:
\begin{equation}
    \beta = \tan\left(\frac{\pi \lambda}{2}\right), \quad R_g = \frac{2^{2\lambda+3}}{\sqrt{\pi}} \frac{\Gamma(\lambda+5/2)}{\Gamma(\lambda+4)}.
\end{equation}

We use two standard wavefunction parameterizations \cite{Kowalski:2006hc}: Gaus-LC and Boosted Gaussian, to evaluate the overlap integral in Eq. (\ref{eq:amppp}). For the Gaus-LC  parameterization, the scalar part of the vector meson wavefunction is assumed to factorize into the transverse size $r$ and the longitudinal momentum fraction $z$. It takes a simplified Gaussian form:
\begin{equation}
    \phi_{\text{Gaus-LC}}(r, z) = N_G z(1-z) \exp\left( -\frac{r^2}{2R_G^2} \right),
    \label{eq:gaus_lc}
\end{equation}
where $R_G$ controls the transverse size of the meson. In contrast, the Boosted Gaussian parameterization is derived by boosting a non-relativistic Gaussian wavefunction of the $q\bar{q}$ pair to the light cone. This approach introduces a correlation between $r$ and $z$ in the exponential term:
\begin{equation}
    \phi_{\text{BG}}(r, z) = N_{\text{BG}} z(1-z) \exp\left( -\frac{m_f^2 \mathcal{R}^2}{8z(1-z)} - \frac{2z(1-z)r^2}{\mathcal{R}^2} \right),
    \label{eq:boosted_gaus}
\end{equation}
where $m_f$ is the quark mass and $\mathcal{R}$ is the variational scale parameter. In both parametrizations, the normalization constants ($N_G, N_{\text{BG}}$) and the size parameters ($R_G, \mathcal{R}$) are constrained by the experimental electronic decay width $\Gamma(V \to e^+ e^-)$ and the wavefunction normalization conditions~\cite{Watt:2007nr}.

\section{PINN Framework and Training Strategy}
\label{sec:method}

\subsection{Network Architecture and Workflow}
The evolution of the color dipole amplitude $N(k, Y)$ is governed by the BK equation. Instead of relying on rigid parametric ansatzes common in phenomenology, we treat the determination of $N$ as a PDE-constrained optimization problem via a fully connected PINN.

We employ a fully connected neural network (FCNN) to approximate the dipole amplitude $N(L, Y)$. The detailed architecture and workflow are illustrated in Fig.~\ref{fig:workflow}. The network consists of:
\begin{itemize}
    \item \textbf{Input Layer}: Accepts the logarithmic transverse momentum $L = \ln(k^2/\Lambda_{\text{QCD}}^2)$ and rapidity $Y = \ln(1/x)$.
    \item \textbf{Hidden Layers}: Four fully connected layers, each containing 64 neurons. We use the hyperbolic tangent ($\tanh$) as the activation function for all hidden layers to ensure the smoothness required for computing high-order derivatives in the PDE loss.
    \item \textbf{Output Layer}: A single neuron with a Sigmoid activation function $\sigma(x) = 1/(1+e^{-x})$, strictly constraining the output amplitude to the physical unitarity limit  $0 \leq N \leq 1$. This is implemented as a necessary numerical regularizer to prevent gradient explosion driven by the non-linear $-N^2$ term during optimization, ensuring stable convergence of the PINN, though it introduces a truncation in the deep infrared regime.
\end{itemize}

\begin{figure*}[t]
\centering
\resizebox{0.9\textwidth}{!}{
\begin{tikzpicture}[
    node distance=1.0cm and 1.0cm,
    font=\sffamily\small,
    block/.style={rectangle, draw=black!60, fill=white, thick, rounded corners=2pt, minimum height=1.2cm, minimum width=2.2cm, align=center, inner sep=5pt},
    input/.style={block, fill=blue!5, draw=blue!40},
    solver/.style={block, fill=red!5, draw=red!40, minimum width=3.5cm},
    output/.style={block, fill=green!5, draw=green!40},
    arrow/.style={-Latex, thick, gray!80},
    label/.style={font=\scriptsize\bfseries, text=gray!60},
    note/.style={font=\scriptsize, text=black!70, align=center}
]

\node[input] (kinematics) {
    \textbf{Kinematics}\\
    $L = \ln(k^2/k_0^2)$\\
    $Y = \bar{\alpha}_s \ln(1/x)$
};

\node[input, below=0.3cm of kinematics] (colloc) {
    \textbf{Collocation}\\
    Sample $(L_c, Y_c)$\\
    Boundary Conds.
};

\node[draw=blue!30, dashed, fit=(kinematics) (colloc), inner sep=5pt, label={[blue!50]above:Input Domain}] (input_group) {};

\node[solver, right=1.5cm of input_group] (pinn) {
    \textbf{BK-PINN Solver}\\
    Ansatz: $N \approx A_0(L)\sigma(f_\theta)$
};

\node[block, draw=orange!40, fill=orange!5, above=0.8cm of pinn, minimum width=4cm] (loss) {
    \textbf{Physics \& Data Loss}\\
    $\mathcal{L} = \lambda_1 \mathcal{L}_{\text{PDE}} + \lambda_2 \mathcal{L}_{\text{Data}} + \lambda_3 \mathcal{L}_{\text{Bound}}$
};

\node[block, draw=gray!40, fill=gray!5, below=0.8cm of pinn, minimum width=3.5cm] (opt) {
    \textbf{Optimization}\\
    Adam + L-BFGS\\
    AD (Automatic Differentiation): $\partial_Y N, \partial_L^2 N$
};

\node[input, above=0.8cm of loss, fill=yellow!5, draw=yellow!60] (data) {
    \textbf{HERA DIS Data}\\
    $F_2(x, Q^2)$ bins\\
    (ZEUS / H1)
};

\node[output, right=1.5cm of pinn] (amplitude) {
    \textbf{Dipole Amplitude}\\
    $N(k, Y) \pm 2\sigma$\\
    (Mean $\pm$ Std)
};

\node[output, right=1.9cm of amplitude] (predict) {
    \textbf{Exclusive Process}\\
    $\gamma^* p \to J/\psi p$\\
    $d\sigma/dt \propto |\mathcal{A}|^2$
};


\draw[arrow] (input_group) -- node[above, font=\scriptsize]{Feed} (pinn);

\draw[arrow] (pinn) -- (opt);
\draw[arrow] (opt.east) to[out=0,in=-45] (pinn.east);

\draw[arrow] (pinn) -- (loss);
\draw[arrow] (loss) -- node[right, font=\scriptsize]{Update $\theta$} (pinn);

\draw[arrow] (data) -- (loss);

\draw[arrow] (pinn) -- node[above, font=\scriptsize]{Predict} (amplitude);

\draw[arrow] (amplitude) -- node[above, font=\scriptsize]{Fourier-Bessel} node[below, font=\scriptsize]{$k \to r$} (predict);

\begin{scope}[on background layer]
    \node[draw=red!20, fill=red!2, rounded corners, fit=(loss) (pinn) (opt), inner sep=8pt, label={[red!40]below:Training Loop}] {};
\end{scope}

\end{tikzpicture}
}
\caption{Workflow of the BK-constrained PINN framework. The training process integrates physical constraints (BK equation) and experimental data (HERA $F_2$) through a composite loss function. The optimized network yields the momentum-space dipole amplitude $N(k,Y)$, which is then transformed to coordinate space to predict exclusive vector meson production cross-sections.}
\label{fig:workflow}
\end{figure*}

\subsection{Teacher-Student Strategy}
We employ a ``Teacher-Student'' training strategy to ensure physical consistency. The total loss $\mathcal{L} = \lambda_1 \mathcal{L}_{\text{PDE}} + \lambda_2 \mathcal{L}_{\text{Data}} + \lambda_3 \mathcal{L}_{\text{Bound}}$ is minimized in two phases (see Fig. \ref{fig:pinn_arch}):
\begin{itemize}
    \item \textbf{Phase I (Teacher)}: The network is trained solely on the PDE loss $\mathcal{L}_{\text{PDE}}$ (Eq.~\ref{eq:BK_diff}). This enforces the BK evolution dynamics, acting as a ``teacher'' to establish the physical manifold of the solution before seeing data.
    \item \textbf{Phase II (Student)}: The data loss $\mathcal{L}_{\text{Data}}$ (from HERA $F_2$) is activated. The network (``student'') refines the solution to match experimental observations while remaining constrained by the physics learned in Phase I.
\end{itemize}

This approach allows the network to approximate the dipole amplitude as a universal function approximator, free from ad-hoc model biases. To bridge the theoretical amplitude with inclusive DIS measurements, we establish a direct link to the structure function $F_2(x, Q^2)$ strictly within momentum space (Eq.~\ref{eq:f2}), thereby bypassing the numerical instabilities associated with Fourier-Bessel transforms. Assuming the proton behaves as a homogeneous ``black disk'' within an effective radius $R_p$ (integrated impact parameter dependence) \cite{Wang:2022jwh,Wang:2023poi}, $R_p$ becomes the sole geometric parameter determined by the inclusive fit.

\subsection{Uncertainty Quantification}
To quantify epistemic uncertainty, we use the Bootstrap Aggregating (Bagging) method \cite{NNPDF:2017mvq,Efron:1979bxm,lakshminarayanan2017simplescalablepredictiveuncertainty}. The HERA $F_2$ dataset is randomly partitioned into a training set ($80\%$) and a testing set ($20\%$). We train an ensemble of $N=20$ independent PINNs on resampled datasets. The final prediction is the ensemble mean, and the uncertainty band corresponds to $\pm 2\sigma$ (standard deviation).

\begin{figure}[t]
\centering
    \begin{tikzpicture}[
        scale=0.85, transform shape,
        node distance=0.6cm and 1.2cm,
        layer/.style={rectangle, draw=black!70, fill=blue!5, thick, minimum height=2.4cm, minimum width=1.0cm, align=center},
        neuron/.style={circle, draw=black, fill=white, minimum size=0.5cm, inner sep=0pt},
        process/.style={rectangle, draw=red!70, fill=red!5, thick, rounded corners, minimum height=0.9cm, minimum width=2.4cm, align=center, font=\sffamily\footnotesize},
        data/.style={rectangle, draw=green!60!black, fill=green!5, thick, dashed, minimum height=0.9cm, minimum width=2.4cm, align=center, font=\sffamily\footnotesize},
        loss/.style={ellipse, draw=orange!80!black, fill=orange!10, thick, minimum height=0.7cm, inner sep=2pt, font=\sffamily\footnotesize, align=center},
        conn/.style={-Latex, thick, gray},
        txt/.style={font=\sffamily\scriptsize}
    ]

    \node[neuron] (I1) at (0, 0.4) {$L$};
    \node[neuron] (I2) at (0, -0.4) {$Y$};
    \node[below=0.15cm of I2, font=\scriptsize] {Input};
    
    \node[layer] (H1) at (1.8, 0) {};
    \node[above=0.05cm of H1, font=\scriptsize] {H1 (64)};
    \foreach \y in {0.9, 0.3, -0.3, -0.9}
        \node[neuron, fill=blue!10, scale=0.8] at (1.8, \y) {};
        
    \node[font=\normalsize] (dots) at (3.0, 0) {$\dots$};
    
    \node[layer] (H4) at (4.2, 0) {};
    \node[above=0.05cm of H4, font=\scriptsize] {H4 (64)};
    \foreach \y in {0.9, 0.3, -0.3, -0.9}
        \node[neuron, fill=blue!10, scale=0.8] at (4.2, \y) {};

    \node[neuron, fill=red!10, minimum size=0.8cm] (O) at (6.0, 0) {$N$};
    \node[below=0.15cm of O, font=\scriptsize] {Sigmoid};

    \draw[conn] (I1) -- (H1); \draw[conn] (I2) -- (H1);
    \draw[conn] (H1) -- (dots); \draw[conn] (dots) -- (H4);
    \draw[conn] (H4) -- (O);

    \node[process, above right=0.3cm and 1.8cm of O] (autodiff) {Auto-Diff\\$\partial N/\partial L, \dots$};
    \node[process, right=0.8cm of autodiff] (bk) {BK Residual\\(PDE)};
    \node[loss, right=0.6cm of bk] (lpde) {$\mathcal{L}_{\text{PDE}}$};
    \node[txt, above=0.05cm of autodiff, red!70!black] {Phase I: Teacher (Physics)};

    \draw[conn, rounded corners] (O) |- (autodiff);
    \draw[conn] (autodiff) -- (bk);
    \draw[conn] (bk) -- (lpde);

    \node[process, below right=0.3cm and 1.8cm of O] (integ) {Momentum Integ.\\$\int dk |\Psi|^2 N/k$};
    \node[process, right=0.8cm of integ] (f2calc) {Structure Func.\\$F_2(x, Q^2)$};
    \node[loss, right=0.6cm of f2calc] (ldata) {$\mathcal{L}_{\text{Data}}$};
    \node[data, below=0.4cm of f2calc] (exp) {HERA Data};
    \node[txt, below=0.05cm of integ, green!60!black] {Phase II: Student (Data)};

    \draw[conn, rounded corners] (O) |- (integ);
    \draw[conn] (integ) -- (f2calc);
    \draw[conn] (f2calc) -- (ldata);
    \draw[conn, dashed] (exp) -- (f2calc);

    \coordinate (mid) at ($(lpde)!0.5!(ldata)$);
    \node[loss, fill=yellow!20, right=1.0cm of mid, minimum width=1.8cm, minimum height=0.9cm] (total) {Total Loss};
    
    \draw[conn] (lpde) -| (total);
    \draw[conn] (ldata) -| (total);

    \end{tikzpicture}
    \caption{Schematic of the BK-PINN architecture and data flow. The training is governed by two loss components: the physics-based $\mathcal{L}_{\text{PDE}}$ derived from the BK equation via automatic differentiation, and the data-driven $\mathcal{L}_{\text{Data}}$ obtained by directly integrating the amplitude with photon wavefunctions.}
    \label{fig:pinn_arch}
\end{figure}

\section{Results and Discussion}
\label{sec:results}

\subsection{Model Validation and Inclusive Fit}
\begin{figure}[h]
    \centering
    \includegraphics[width=0.49\linewidth]{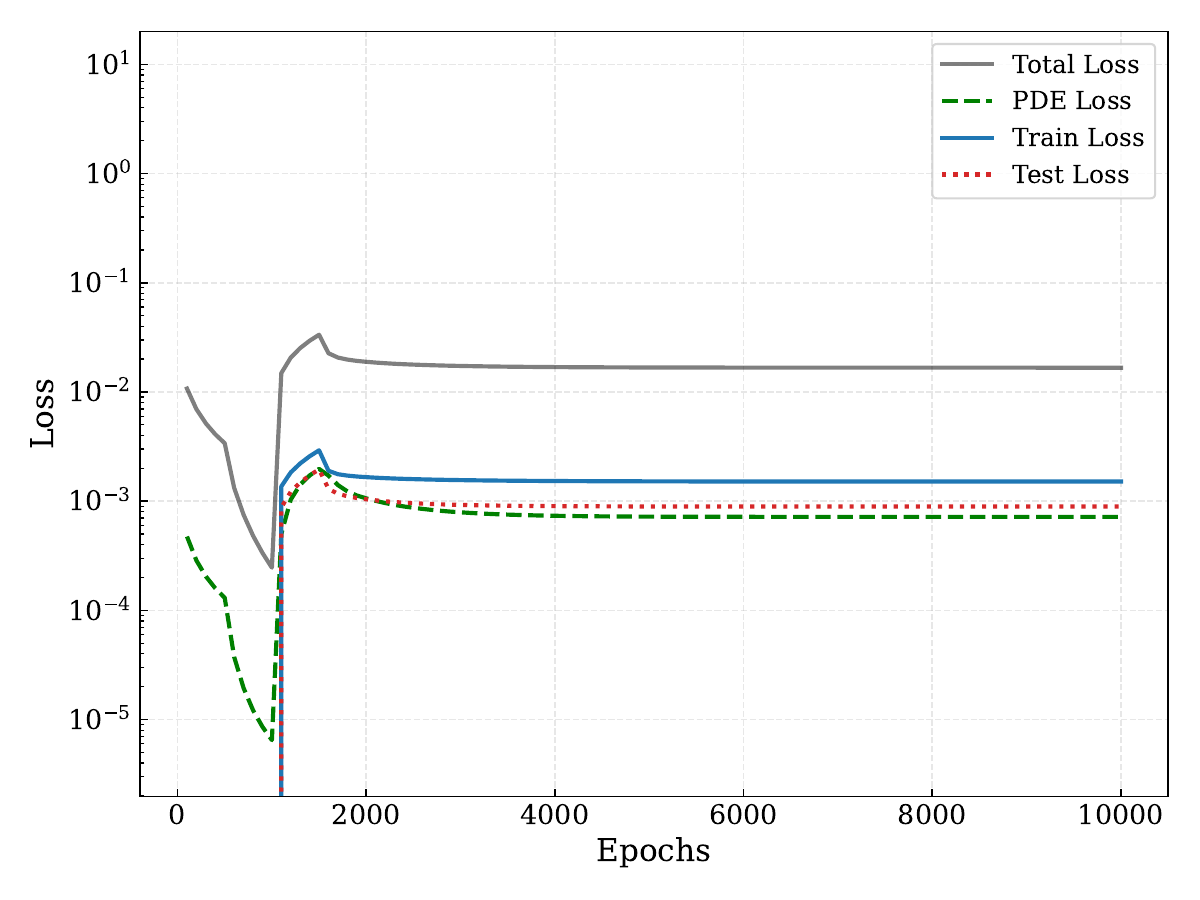}
    \caption{Training dynamics showing the total loss, PDE loss (Teacher), and Data loss (Student) for both training and testing sets.}
    \label{fig:loss}
\end{figure}

We first assess the convergence of the PINN training. Figure~\ref{fig:loss} shows the loss history, demonstrating an initial drop in the PDE loss (Teacher phase) followed by the synchronous convergence of data losses (Student phase).

\begin{figure}[t]
    \centering
    \includegraphics[width=0.99\linewidth]{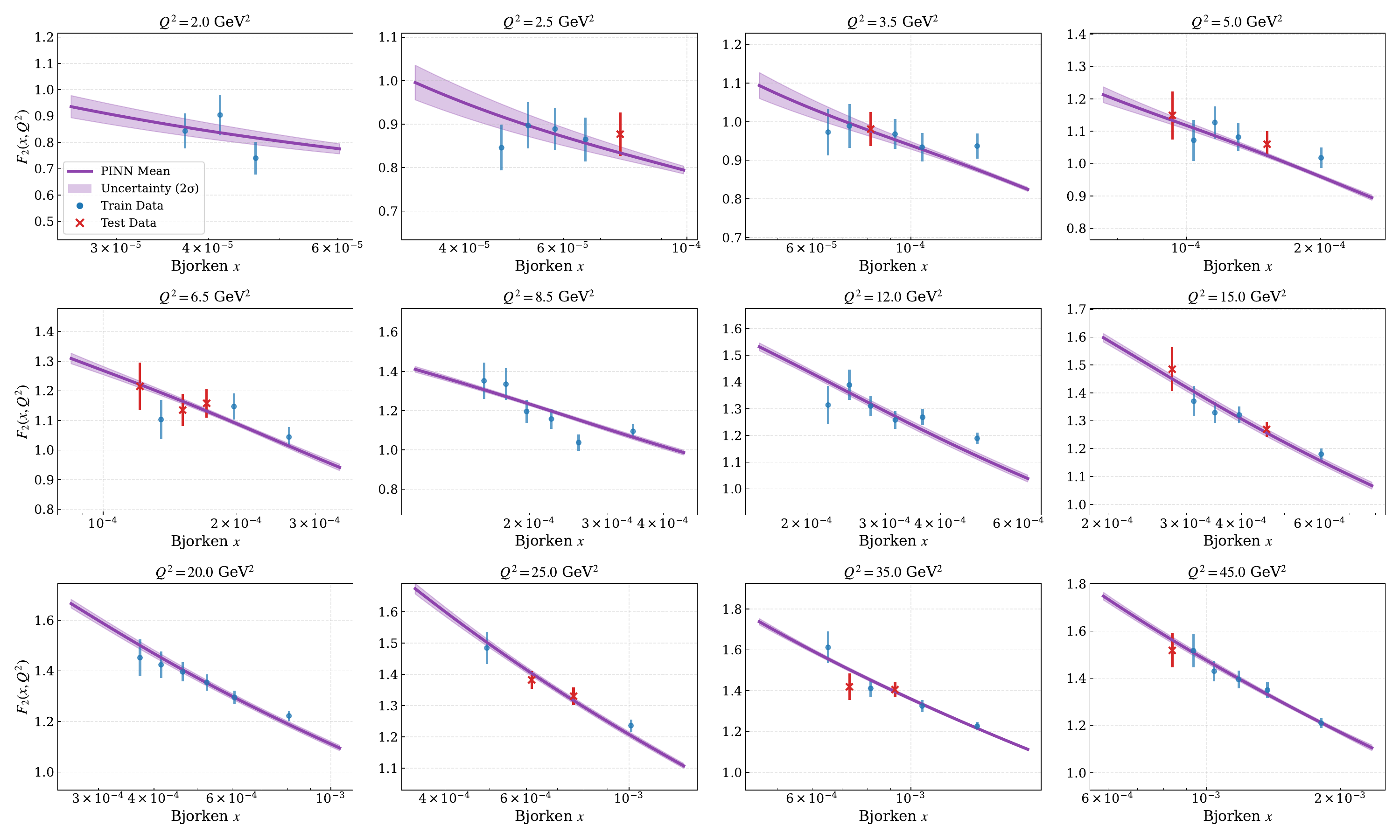}
    \caption{PINN fit to inclusive $F_2$ data \cite{H1:2009pze,H1:2013ktq}. The shaded bands represent the $2\sigma$ uncertainty derived from the ensemble.}
    \label{fig:f2}
\end{figure}

The resulting global fit to the inclusive HERA $F_2$ data \cite{H1:2009pze,H1:2013ktq} is visualized in Fig.~\ref{fig:f2}. The PINN model accurately captures the kinematic dependence of the structure function across the entire $(x, Q^2)$ plane covered by HERA. The shaded bands indicate the epistemic uncertainty derived from the ensemble, showing that the network is robust even in kinematic regions with sparse data. Notably, the uncertainty bands appear to narrow towards higher $Q^2$ (corresponding to the small-$r$ or high-$k$ perturbative regime). While inclusive DIS data typically provide weak constraints on the dipole amplitude at extremely small distances, this reduction in epistemic uncertainty highlights a unique advantage of the PINN framework. In the high-$k$ region, the network's solution manifold is strictly locked by the perturbative linear tail (BFKL dynamics) embedded within the governing equation. Consequently, the rigorous PDE loss ($\mathcal{L}_{\text{PDE}}$) effectively acts as a theoretical regularizer, suppressing the variance of the ensemble and yielding highly robust predictions even where experimental data are sparse.

\begin{figure}[h]
    \centering
    \begin{minipage}{0.48\textwidth}
        \centering
        \includegraphics[width=\linewidth]{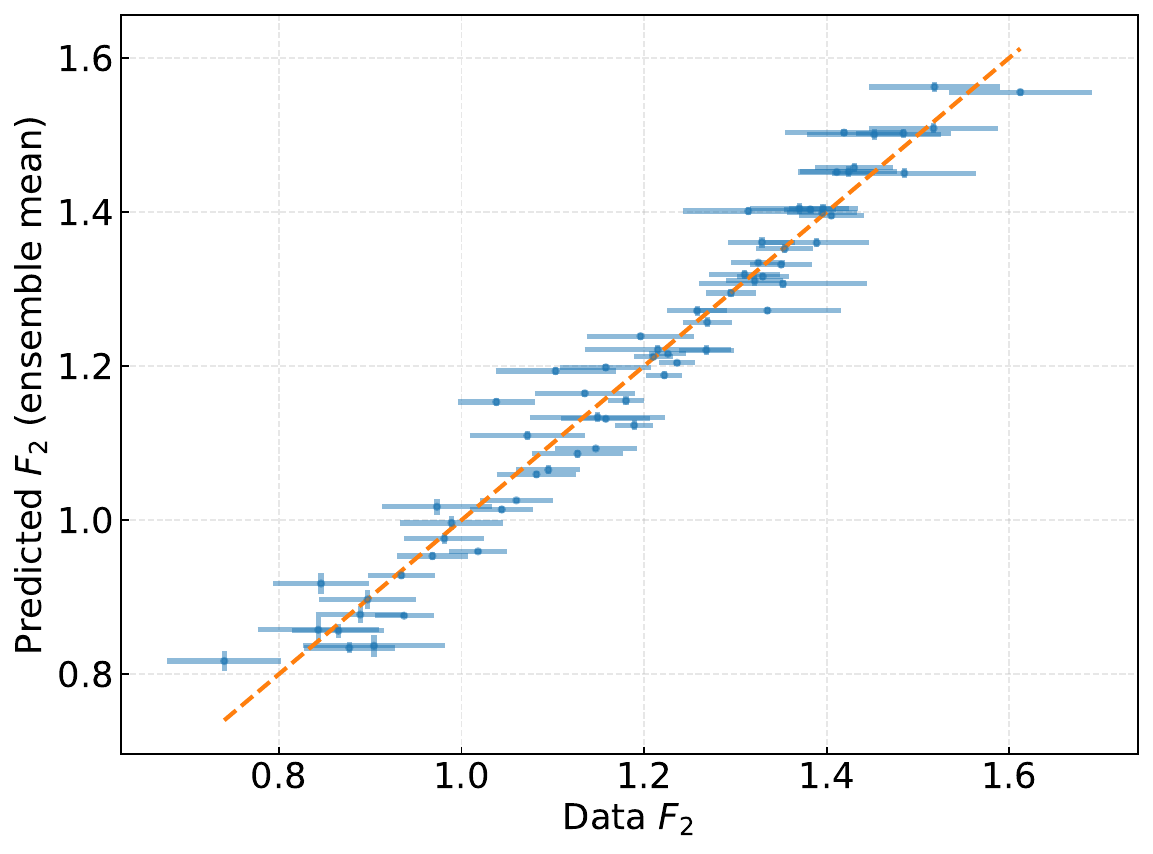}
    \end{minipage}\hfill
    \begin{minipage}{0.48\textwidth}
        \centering
        \includegraphics[width=\linewidth]{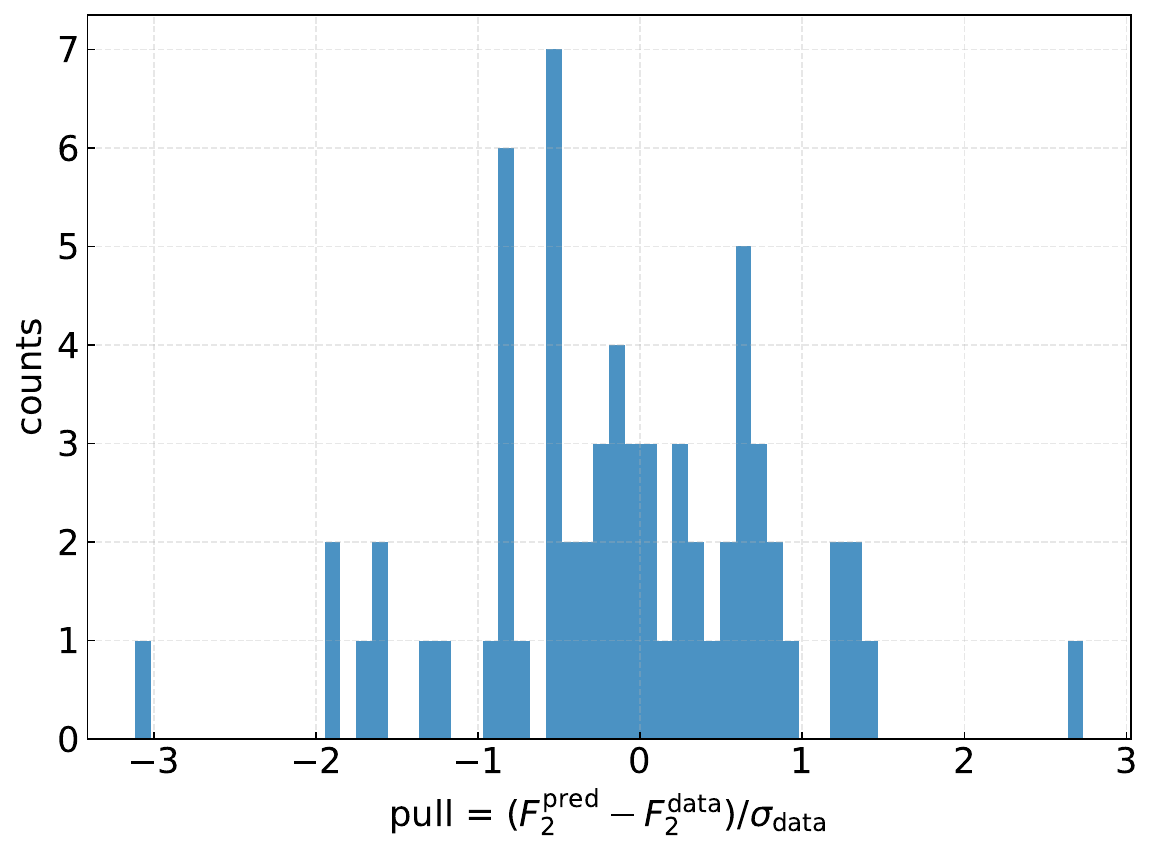}
    \end{minipage}
    \caption{Left: Parity plot of predicted vs. experimental $F_2$. Right: Pull distribution of the residuals.}
    \label{fig:stats}
\end{figure}

To quantitatively assess the statistical quality of this fit, we analyze the distribution of residuals in Fig.~\ref{fig:stats}. The parity plot (Left) shows that the predicted values cluster tightly around the diagonal $y=x$ line, indicating no systematic bias. The pull distribution (Right), defined as $(F_2^{\text{pred}} - F_2^{\text{data}}) / \sigma_{\text{data}}$, approximates a standard normal Gaussian ($\mu \approx 0, \sigma \approx 1$), confirming that the model provides an unbiased description within experimental uncertainties. Quantitatively, the ensemble mean prediction yields a global $\chi^2 / \text{dof} \approx 0.972$ for the inclusive HERA $F_2$ data (where the degrees of freedom are defined by the number of data points minus the 5 extracted physical parameters). This value, being close to unity, demonstrates the high statistical fidelity of the PINN solution.

\subsection{Physical Parameters and Emergent Dynamics}
A key advantage of the PINN framework is its ability to extract the underlying parameters of the governing equation from data. We analyze the ensemble of trained networks to infer the effective coefficients of the BK equation in the diffusion approximation. Table~\ref{tab:params} summarizes the extracted parameters. Using these extracted coefficients, the effective saturation exponent can be derived according to the traveling-wave physics as $\lambda_s = (A_1 + 2\sqrt{A_0 A_2}) \bar{\alpha}_s \approx 0.239 \pm 0.010$, which is in remarkable agreement with phenomenological extractions from HERA data (e.g., $\lambda \approx 0.288$ in the GBW model). The extracted proton radius parameter $R_p \approx 5.46$ GeV$^{-1}$ (approx. 1.07 fm) reflects the effective gluonic radius.

\begin{table}[h]
    \centering
    \caption{Mean values and standard deviations of the physical parameters extracted from the PINN ensemble. The drift coefficient $A_1$ is learned by the network to accommodate the specific rapidity frame and potential higher-order corrections.}
    \begin{tabular}{lccc}
        \toprule
        Parameter & Symbol & Mean & Std \\
        \midrule
        Growth Rate & $A_0$ & 0.661 & 0.004 \\
        Drift Rate  & $A_1$ & 0.414 & 0.013 \\
        Diffusion Rate & $A_2$ & 0.266 & 0.033 \\
        Effective Proton Radius & $R_p$ [GeV$^{-1}$] & 5.4560 & 0.0004 \\
        \bottomrule
    \end{tabular}
    \label{tab:params}
\end{table}

\begin{figure}[tbp]
    \centering
    \begin{minipage}{0.48\textwidth}
        \centering
        \includegraphics[width=\linewidth]{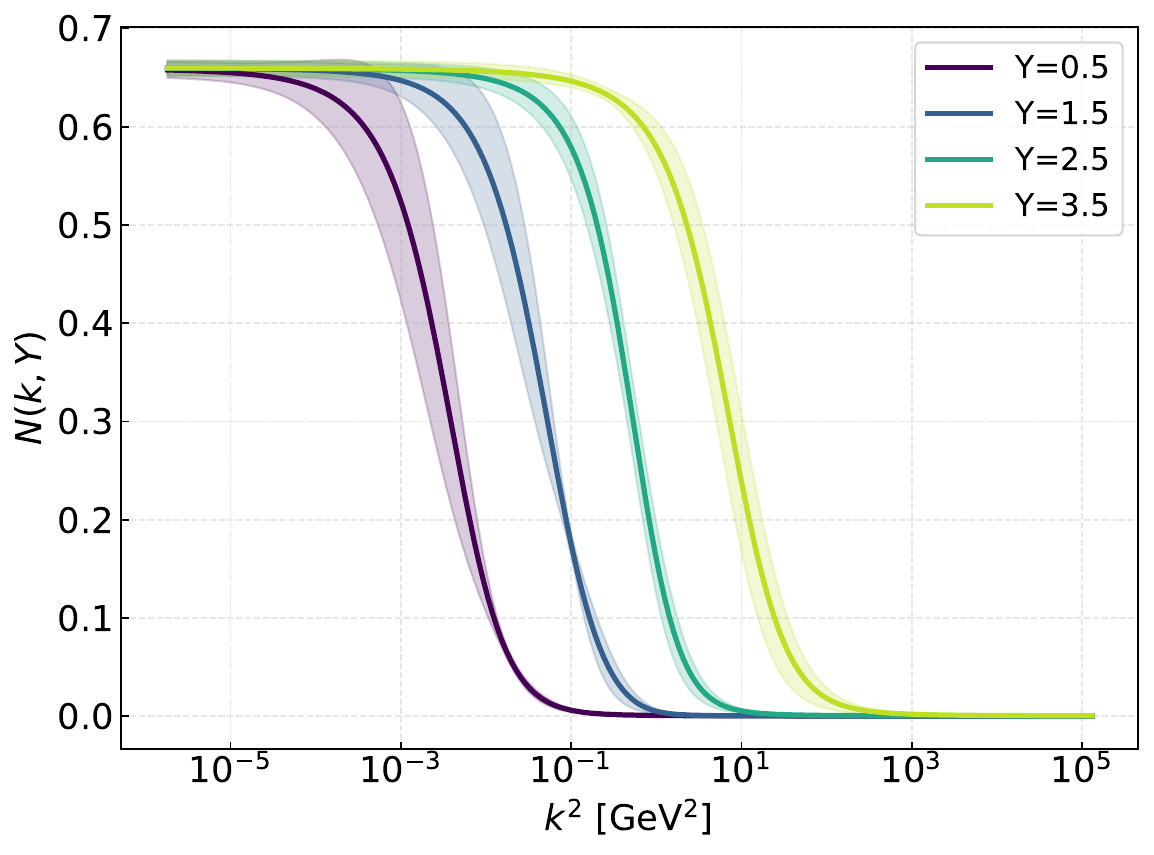}
    \end{minipage}\hfill
    \begin{minipage}{0.48\textwidth}
        \centering
        \includegraphics[width=\linewidth]{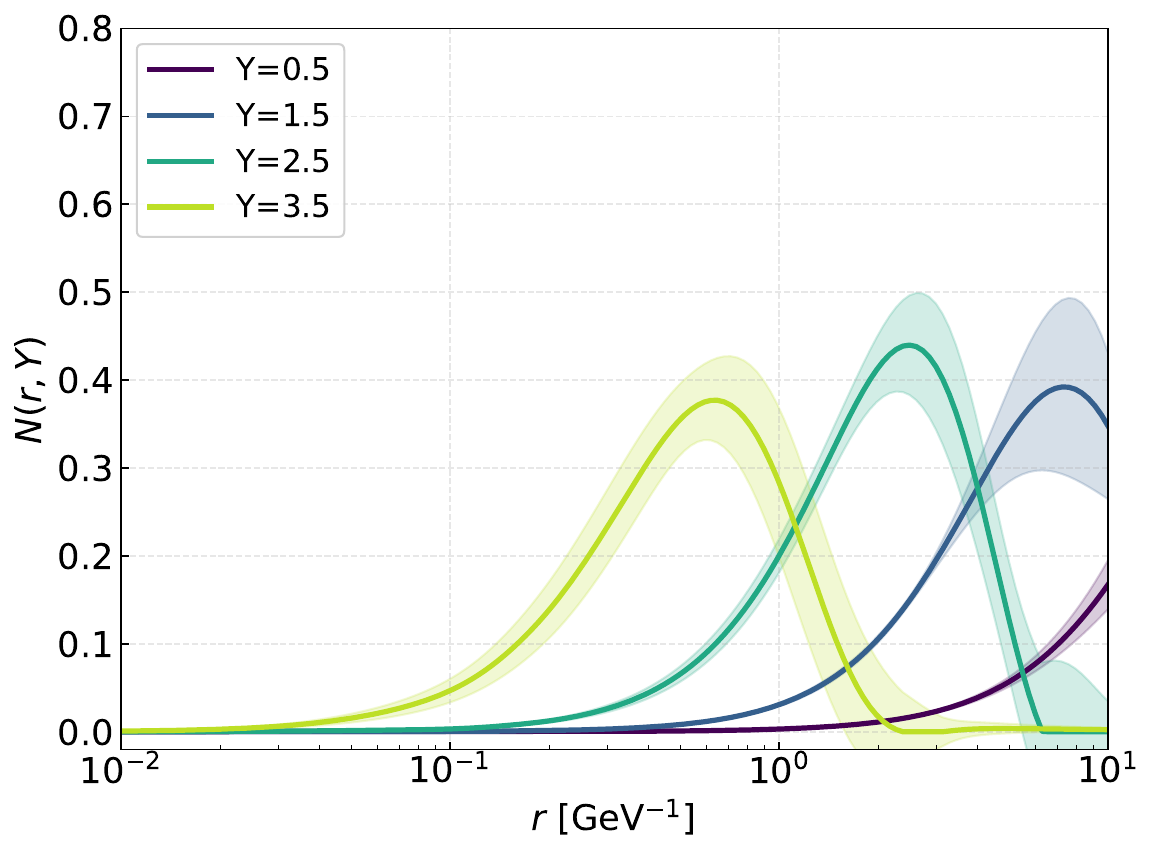}
    \end{minipage}
    \caption{Evolution of the extracted dipole amplitude with rapidity $Y=0.5, 1.5, 2.5, 3.5$ (from left to right). Left: Momentum-space amplitude $N(k, Y)$. Right: Coordinate-space amplitude $N(r, Y)$ obtained via the Hankel transform. The shaded bands represent the $2\sigma$ uncertainty derived from the ensemble. The wavefront propagation illustrates the dynamic generation of the saturation scale $Q_s(Y)$.}
    \label{fig:amplitude}
\end{figure}

As shown in Fig.~\ref{fig:amplitude}, the trained PINN solution spontaneously exhibits geometric scaling \cite{Munier:2003vc}. In the momentum space (Left panel), the saturation wavefront propagates towards higher transverse momenta as rapidity increases, a feature that emerges naturally from the interplay between linear growth and non-linear damping terms learned by the network. To further scrutinize the physical consistency, we present the dipole amplitude in coordinate space, $N(r, Y) = r^2 \int_0^\infty k \, dk \, J_0(kr) N(k, Y)$, in the right panel of Fig.~\ref{fig:amplitude}. The coordinate-space amplitude monotonically increases and captures the onset of saturation in the physically relevant perturbative and transition regions ($r \lesssim 2 \text{ GeV}^{-1}$). 

It is noteworthy that the amplitude exhibits a decay at extremely large transverse distances ($r > 3 \text{ GeV}^{-1}$). This large-$r$ decay is a direct mathematical consequence of our numerical regularization in momentum space. In the saturation regime, the PINN output is effectively bounded by the learned growth parameter $A_0$. Because the network actively bounds $N(k,Y)$ in the infrared to maintain numerical stability, it lacks the $1/k^2$ pole required for $N(r) \to 1$ as $r \to \infty$. Consequently, according to the Riemann-Lebesgue lemma, its Hankel transform must mathematically decay to zero in the deep infrared limit. Crucially, this decay has negligible impact on our phenomenological results. For exclusive $J/\psi$ production, the massive charm quark ($m_c \approx 1.4$ GeV) ensures that the vector meson wavefunction acts as a natural infrared filter, exponentially suppressing the overlap integral for $r \gtrsim 1.5 \text{ GeV}^{-1}$. Within this physical domain probed by HERA, our PINN provides a highly accurate and robust effective amplitude, as evidenced by the zero-parameter predictions in the following section.

\subsection{Predictions for Exclusive Processes}
A key test of a robust saturation framework is its process-independence. We subject our PINN solution, trained \textit{solely} on inclusive $F_2$ data, to a rigorous consistency check by predicting the exclusive vector meson photoproduction $\gamma p \to V p$. It is important to clarify the origin of all inputs used in the exclusive calculation. The geometric parameter $R_p$ and the dipole amplitude $N(k, Y)$ (as well as derived quantities like the effective power $\lambda$) are determined by our PINN fit to inclusive data. Conversely, the diffractive slope $B_D$ and the wavefunction parameters are empirical inputs taken from the literature. Thus, applying the PINN amplitude to exclusive processes constitutes a parameter-free prediction of the saturation dynamics, without additional tuning of the dipole amplitude itself.

\begin{figure}[h]
	\centering
	\includegraphics[width=0.49\columnwidth]{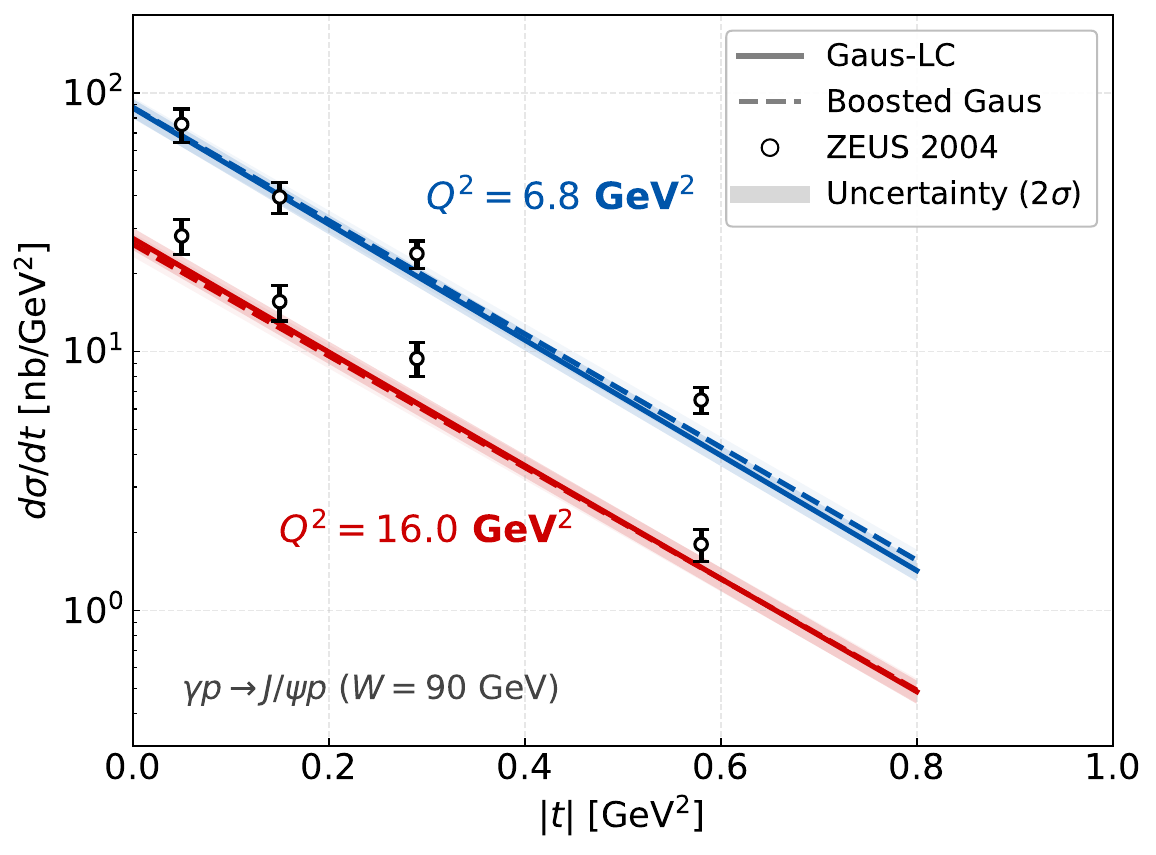} 
    \includegraphics[width=0.49\columnwidth]{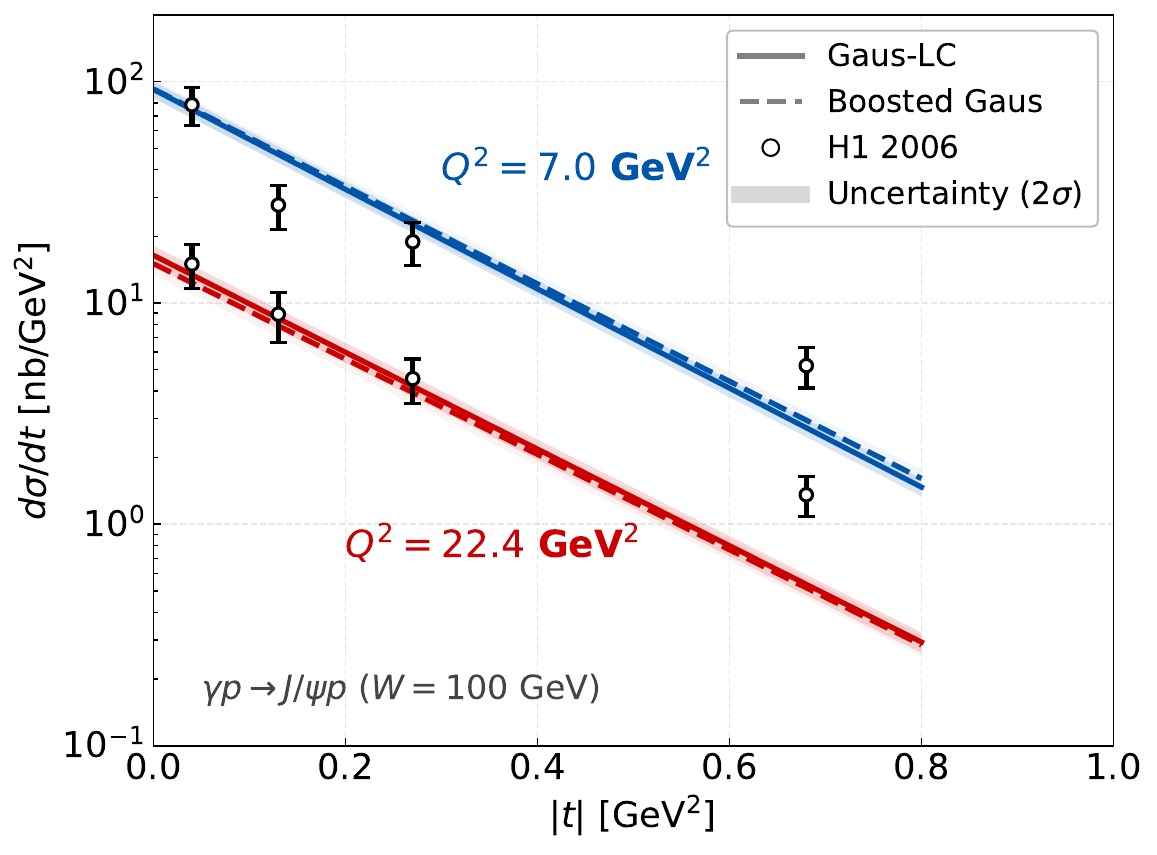}
	\caption{Predictions without additional tuning of the dipole amplitude for exclusive $J/\psi$ photoproduction differential cross-sections. Left: Comparison with ZEUS data at $W=90$ GeV \cite{ZEUS:2004yeh}. Right: Comparison with H1 data at $W=100$ GeV \cite{H1:2005dtp}. The solid and dashed lines represent predictions using Gaus-LC and Boosted Gaussian wavefunctions, respectively.}
	\label{fig:diffractive}
\end{figure}

Figure~\ref{fig:diffractive} presents the comparison between our predictions and HERA measurements for $J/\psi$ production \cite{ZEUS:2004yeh,H1:2005dtp}. The agreement is striking: given the empirical diffractive slope $B_D$, the PINN-derived amplitude accurately reproduces the absolute magnitude of the differential cross-section across a wide range of photon virtualities ($6.8 \le Q^2 \le 22.4$ GeV$^2$). Furthermore, the effective proton radius $R_p \approx 5.46 \text{ GeV}^{-1} \approx 1.07$ fm extracted solely from our inclusive fit is physically consistent with the spatial imaging probed by diffraction (where $B_D \approx 4.7 \text{ GeV}^{-2}$ corresponds to a compatible transverse gluonic size). This agreement demonstrates that a dipole amplitude constrained by BK dynamics and fitted solely to inclusive data can successfully describe exclusive observables without retuning, supporting the underlying process-independence of the saturation mechanism.

\begin{figure}[h]
    \centering
    \includegraphics[width=0.49\linewidth]{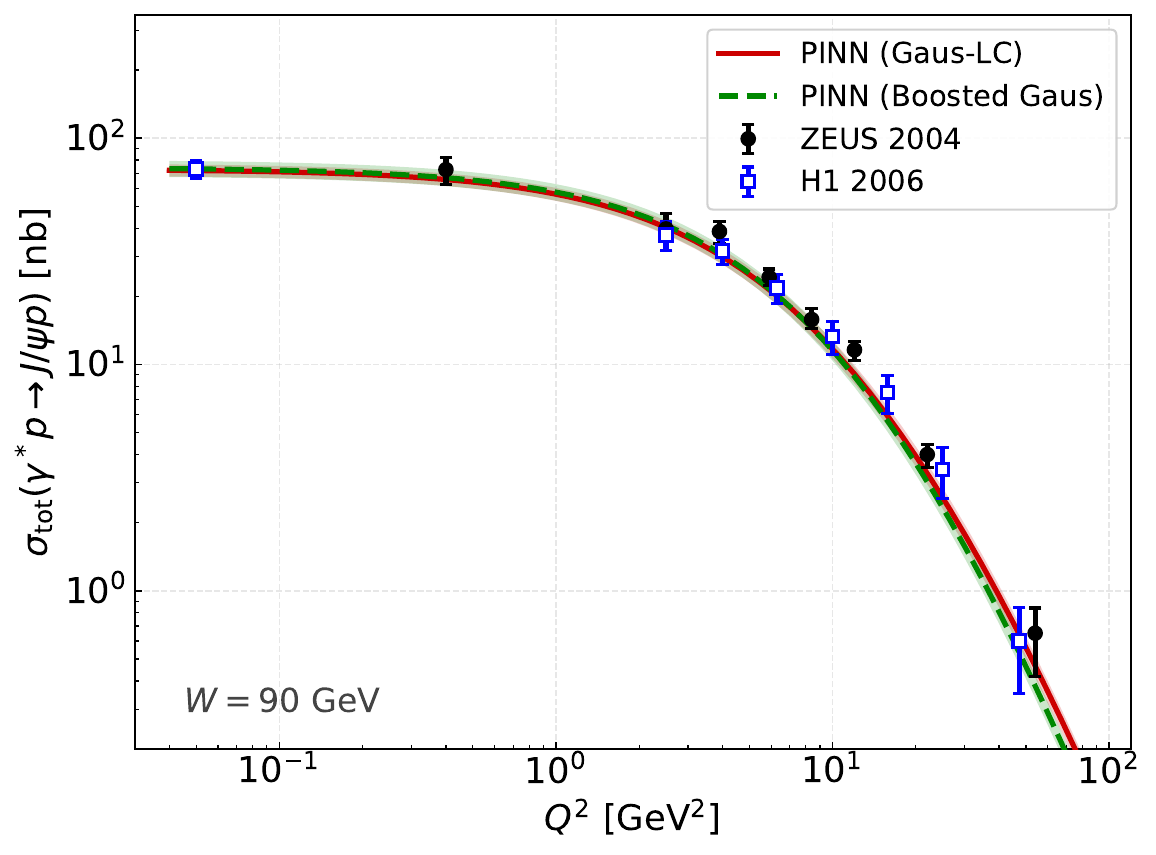}
    \caption{Total cross section $\sigma_{\text{tot}}$ for exclusive $J/\psi$ photoproduction as a function of $Q^2$ at fixed center-of-mass energy $W=90$ GeV. The PINN predictions are compared with experimental data from ZEUS 2004~\cite{ZEUS:2004yeh} and H1 2006~\cite{H1:2005dtp}.}
    \label{fig:total_xs}
\end{figure}

To further verify the predictive power, we calculate the total cross-section for exclusive $J/\psi$ photoproduction as a function of $Q^2$, shown in Fig.~\ref{fig:total_xs}. The model successfully reproduces the $Q^2$ dependence over three orders of magnitude ($0.05 \lesssim Q^2 \lesssim 100$ GeV$^2$), further validating the universality of the saturation mechanism captured by the PINN.

\section{Conclusion}
\label{sec:conclusion}

In summary, we have established a unified description of high-energy scattering by integrating the physics-based momentum-space BK evolution dynamics into a Physics-Informed Neural Network. This framework effectively bridges the gap between perturbative QCD evolution and non-perturbative hadronic structure, resolving the long-standing tension between inclusive and diffractive measurements without ad-hoc geometric adjustments. Quantitatively, the PINN model achieves a high-fidelity global fit to HERA $F_2$ data, yielding a competitive $\chi^2/\text{dof} \approx 0.972$. We have demonstrated that the extracted dipole amplitude maintains physical consistency across both momentum and coordinate spaces, with the emergent geometric scaling and saturation wavefronts arising naturally from the governing evolution equation.  Strikingly, the parameter-free prediction of the saturation dynamics and kinematic dependence of exclusive $J/\psi$ photoproduction cross-sections supports the process-independence of the gluon saturation scale.

 While the current results are compelling, the framework opens clear avenues for systematic refinement. The quantitative limitations of the fixed-coupling diffusion approximation restrict the current applicability to a specific small-$x$ and moderate $Q^2$ kinematic window; future iterations will incorporate full running coupling BK (rcBK) dynamics without diffusion truncation and full next-to-leading order (NLO) corrections and explicit impact-parameter dependence to extend precision across a broader $Q^2$ range. Furthermore, the integration of advanced machine learning architectures, such as Bayesian Neural Networks (BNNs) \cite{neal2012bayesian,Jospin_2022} for rigorous uncertainty quantification and Operator Learning for parameter-dependent solutions, will be essential for maximizing the discovery potential of future high-precision data from the Electron-Ion Collider (EIC) \cite{Accardi:2012qut,AbdulKhalek:2021gbh}. Ultimately, this work demonstrates that physics-informed deep learning is not merely a fitting tool, but a transformative paradigm for uncovering the fundamental laws of strong interaction.

\begin{acknowledgments}
This work has been supported by the National Natural Science Foundation of China (Grant No. 12547118), the Research Program of State Key Laboratory of Heavy Ion Science and Technology, Institute of Modern Physics, Chinese Academy of Sciences (Grant No. HIST2025CS08), and the National Key R$\&$D Program of China (Grant No. 2024YFE0109800 and 2024YFE0109802).
\end{acknowledgments}
\section*{Data and Code Availability}
The source code for the PINN-BK framework, the trained neural network ensemble, and the scripts used to generate the figures in this study are publicly available at the following GitHub repository: \url{https://github.com/wadeykou/PINN-BK-sol/tree/main}.
\bibliography{refs}

\end{document}